\documentclass[12pt,preprint]{aastex}

\usepackage{emulateapj5}

\usepackage{epsfig}
                                                                                
\def\vkm{km s$^{-1}$}

\def\degree{$^\circ$}
\def\degreeqa#1#2{#1^{\circ}_{^\textrm{.}}#2}
\def\arcs#1{$#1''$}
\def\arcsa#1#2{$#1^{\prime\prime}_{^\textrm{.}}#2$}

\def\solarmass{$M_\odot$}

\def\mJyb{mJy beam$^{-1}$}

\def\cmc{cm$^{-3}$}
\def\cms{cm$^{-2}$}

\def\ra#1#2#3#4{#1^\mathrm{h} #2^\mathrm{m} #3^\mathrm{s}_{^\textrm{.}} #4}
\def\dec#1#2#3#4{#1\degr #2\arcmin #3^{\prime\prime}_{^\textrm{.}}#4}

\def\mH{m_\textrm{\scriptsize H}}
\def\nH{n_\textrm{\scriptsize H}}

\def\SOt{$N_J=8_9-7_8$}
\def\H2{H$_2$}
\def\mH2{m_{\textrm{\scriptsize H}_2}}
\def\nH2{n_{\textrm{\scriptsize H}_2}}
\def\N2HP{N$_2$H$^+$}

\def\NH3{NH$_3$}

\def\SOt{$N_J=8_9-7_8$}

\def\putfig#1#2#3{\epsfig{scale=#1,angle=#2,figure=#3}}
\def\putfiga#1#2#3{}
\def\leftblank#1{}

\begin{document}

\title{Magnetic field structure in the Flattened Envelope and Jet in the
young protostellar system HH 211} \author{Chin-Fei Lee\altaffilmark{1},
Ramprasad Rao\altaffilmark{1},
Tao-Chung Ching\altaffilmark{2},
Shih-Ping Lai\altaffilmark{2},
Naomi Hirano\altaffilmark{1},
Paul T.P. Ho\altaffilmark{1,3}, and
Hsiang-Chih Hwang\altaffilmark{1}
}
\altaffiltext{1}{Academia Sinica Institute of Astronomy and Astrophysics,
P.O. Box 23-141, Taipei 106, Taiwan; cflee@asiaa.sinica.edu.tw}
\altaffiltext{2}{Institute of Astronomy and Department of Physics, National
Tsing Hua University, Hsinchu, Taiwan}
\altaffiltext{3}{Harvard-Smithsonian Center for Astrophysics, 60 Garden
Street, Cambridge, MA 02138}

\begin{abstract}

HH 211 is a young Class 0 protostellar system, with a flattened envelope, a
possible rotating disk, and a collimated jet.  We have mapped it with the
Submillimeter Array in 341.6 GHz continuum and SiO J=8-7 at $\sim$
\arcsa{0}{6} resolution.  The continuum traces the thermal dust emission in
the flattened envelope and the possible disk.  Linear polarization is
detected in the continuum in the flattened envelope.  The field lines
implied from the polarization have different orientations, but they are not
incompatible with current gravitational collapse models, which predict
different orientation depending on the region/distance.  Also, we might have
detected for the first time polarized SiO line emission in the jet due to
the Goldreich-Kylafis effect.  Observations at higher sensitivity are needed
to determine the field morphology in the jet.

\end{abstract}


\section{Introduction}

Magnetic field has important effects in low-mass star formation. It can
launch a collimated jet but may suppress the formation of a rotationally
supported (accretion) disk (RSD) in the early phase of star formation.  In
current theory, gravitational collapse of a rotating magnetized cloud core
produces a flattened envelope and an hourglass field morphology around a
central source (i.e., protostar) \cite[see, e.g.,][]{Allen2003}.  The
flattened envelope is not rotationally supported and is thus called pseudodisk. 
In the flattened envelope, rotation velocity increases toward the center,
generating a toroidal field in the inner part \citep{Allen2003,Kataoka2012}. 
However, the pinched geometry of the magnetic field in the flattened
envelope generates a magnetic braking, suppressing the formation of a RSD
around the center \citep{Allen2003}.  One possible solution is to have a
misalignment between rotation axis and magnetic field axis
\citep{Joos2012,Li2013}.

Observations of linearly polarized thermal dust emission toward low-mass
Class 0 systems have revealed magnetic field morphologies of the envelope in
support of the above theoretical picture.  They have revealed an hourglass
field morphology in, e.g., NGC 1333 IRAS 4A \citep{Girart2006} and L 1157
\citep{Stephens2013}.  They have also showed a tentative detection of
toroidal fields near the center in, e.g., NGC 1333 IRAS 4A
\citep{Crutcher2012} and IRAS 16293-2422 B \citep{Rao2014}.  Moreover,
misalignment between rotation axis and magnetic field axis could be common,
as suggested in \citet{Hull2013}.  However, further observations are still
needed to confirm the field morphology near the center and the misalignment.

On the other hand, the field morphology in protostellar jet is poorly
determined due to a lack of polarization detection.  In current
jet-launching models \citep{Shu2000,Pudritz2007}, poloidal field is needed
to launch the jet and toroidal field is needed to collimate the jet. 
Therefore, the jet is expected to be magnetized with a helical field.  For
high-mass systems, the jet can emit synchrotron radiation, allowing
us to map the field morphology with polarization observation in synchrotron
continuum \citep{Carrasco2010}.  For low-mass systems, the jet can
emit molecular line emission \citep{Lee2007,Hirano2010}.  Line polarization
has been detected in molecular outflows in low-mass systems, e.g., NGC 1333
IRAS 4A \citep{Girart1999} and high-mass systems, e.g., DR 21
\citep{Lai2003}.  It is attributed to the Goldreich-Kylafis (GK) effect
\citep{Goldreich1981,Goldreich1982}, and thus can be used to infer the field
morphology in the molecular outflows.  It can also be used to infer the
field morphology in the jet.

HH 211 is a nearby (280 pc) low-mass system, in which a flattened envelope
has been detected, a RSD has been claimed, and a collimated jet has been
detected in SiO and CO down to the launching point inside an outflow cavity
(Gueth et al.  1999; Lee et al.  2009, hereafter \citet{Lee2009}).  It is young in
the Class 0 phase, with significant material still in the flattened envelope
and a mass of only $\sim$ 0.06 \solarmass{} for the central source.  The jet
is almost in the plane of the sky and the flattened envelope is almost
edge-on, providing the best view for the fields.  Here, we report detections
of thermal dust polarization in the flattened envelope and line polarization
in the jet, and discuss possible field morphologies in this system.

\section{Observations}

Polarization observation toward HH 211 was carried out with the
Submillimeter Array (SMA)\footnote{The Submillimeter Array is a joint
project between the Smithsonian Astrophysical Observatory and the Academia
Sinica Institute of Astronomy and Astrophysics, and is funded by the
Smithsonian Institution and the Academia Sinica} \citep{Ho2004} on 2013
November 28 in the extended configuration, using a dual-receiver mode with
the 345 GHz and 400 GHz receivers.  SiO ($J=8-7$), CO ($J=3-2$), and SO
(\SOt{}) lines were observed simultaneously with continuum.  Here, we only
present the results in continuum and SiO that show polarization detection. 
The receivers have two sidebands, lower and upper, covering the frequency
range from 335.6 to 337.6 GHz and from 345.6 to 347.6 GHz, respectively. 
Combining the line-free portions of the two sidebands results in a total
bandwidth of $\sim$ 3.7 GHz centered at $\sim$ 341.6 GHz for the continuum. 
The baselines of the array have projected lengths ranging from $\sim$ 27 to
224 m.  The primary beam has a size of $\sim$ \arcs{35}.  The correlator was
set up to have a velocity resolution of $\sim$ 1.4 \vkm{}.

Quasar 3c84 was used as a gain, a bandpass, and a polarization calibrator. 
Its fractional linear polarization was determined to be $\sim$ 0.1\%,
which is the limit of detectability with the SMA. The absolute flux scale
was determined by using the 3c84 flux (6.3 Jy) in the SMA database, with a
flux uncertainty of $\lesssim$ 15\%.  The visibility data were reduced, with
the bandpass and gain calibrations done with the MIR package, and the
polarization and flux calibration done with the MIRIAD package.  The antenna
leakages were found to be up to $\sim$ 3\% except for antenna 7 which is
$\sim$ 7\%.  The calibrated visibility data were imaged with the MIRIAD
package.  With natural weighting, the synthesized beam has a size of
\arcsa{0}{70}$\times$\arcsa{0}{51} at a position angle (P.A.) of $\sim$
$-$88\degree{}.  The rms noise levels in different maps are given later in
figure captions.

\section{Flattened Envelope}

At 341.6 GHz, the continuum map shows a flattened structure within a radius
of $\sim$ \arcs{1} of the central source, roughly perpendicular to the jet
axis (see Fig.  \ref{fig:cont_pol}).  The source is assumed to be at the
peak position of the continuum, which is
$\alpha_{(2000)}=\ra{03}{43}{56}{806}$,
$\delta_{(2000)}=\dec{32}{00}{50}{21}$, consistent with the position found
at higher resolution \citep{Lee2009}.  The jet is known to be bended and
wiggling, with a mean axis having a P.A.  of 114\degree{}$\pm$4\degree{}
\citep{Gueth1999,Lee2007,Lee2009}.  The continuum has a total flux density
of 220$\pm$30 mJy, consistent with the value found before.  The center is
unresolved with a peak flux density of 112 \mJyb{}.  A fit to the
visibility amplitude versus $uv$-distance plot of the continuum requires
two circular Gaussian components; an extended component with a size of
$\sim$ \arcsa{1}{8} and a compact component with a size of $\sim$
\arcsa{0}{3}.  Thus, as discussed before, the continuum traces the thermal
dust emission from a flattened envelope and a possible compact disk around
the source \citep{Lee2007,Lee2009}.  In \citet{Lee2009}, a faint secondary
source was detected at $\sim$ \arcsa{0}{3} to the southwest of the central
source at higher resolution.  However, this source can not be resolved here. 
The envelope is slightly asymmetric, extending further to the SW than to the
NE (probably partly because of the secondary source) and further to the SE
than to the NW.

\subsection{Dust Polarization and Field Morphology}

\begin{figure} [!hbp]
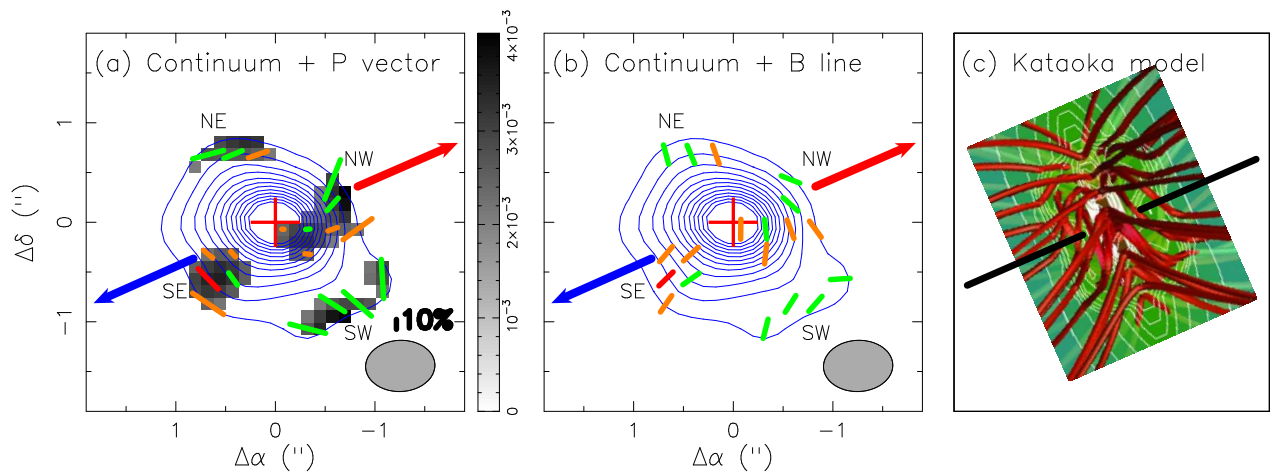

\centering
\putfig{0.72}{270}{f1.ps} %cont_pol0.ps}
\figcaption[]
{Continuum intensity, polarization intensity and vectors, field orientation,
and model field morphology.  In (a) and (b), the blue contours show the
continuum intensity map.  The rms noise level is $\sigma \sim$ 1.2 \mJyb{}.  
The peak has a flux density of 112 \mJyb{}.
The contours start from 5$\sigma$ with a step of 6$\sigma$. The
cross marks the source position.  The blue and red arrows show the
approaching and receding sides of the mean jet axis, respectively.    In (a), the gray-scale
image shows the polarization intensity greater than 2$\sigma$ detection,
where the rms noise level $\sigma=1.15$ \mJyb{}.  Line segments show the polarization
vectors for the detection with 2 to 3.5$\sigma$.
(orange for 2-2.5$\sigma$, green for 2.5-3$\sigma$, and red for
3-3.5$\sigma$).  Polarization degree is indicated by the length of the
vector.  In (b) Line segments show the magnetic field orientations.  (c)
shows the model density (white contours and color image) and field (red lines)
morphology adopted from \citet{Kataoka2012}, with the
field axis (black lines) aligned with the jet axis of HH 211.
\label{fig:cont_pol} } \end{figure}

Linear polarization is detected in the thermal dust emission in the SE, NE,
SW, and NW parts of the envelope, as shown by the polarization
intensity map and vectors in Figure \ref{fig:cont_pol}a.  The angles of the
polarization vectors have an uncertainty of $\sim$ 13\degree{}.  The degree
of polarization reaches $\sim$ 30\% at the edges of the envelope, as found
in other systems e.g., NGC 1333 IRAS 4A \citep{Girart2006}.  In the SE part,
the detection of the polarized emission has a peak of more than 3$\sigma$
and is thus real.  In other parts, the detections are 2 to 3$\sigma$ and
thus should be considered as tentative.

Figure \ref{fig:cont_pol}b shows the magnetic field lines in the envelope
obtained by rotating the polarization vectors by 90\degree{}.  The field
morphology is complicated because the field lines have different
orientations in different regions.  Along the major axis, the SW field lines
are roughly aligned with the jet axis slightly pinched toward the
source. However, the NE field lines, which are closer to the central source
than the SW field lines, are almost perpendicular to the jet axis.
Along the minor axis, the NW field lines appear to be roughly
perpendicular to the jet axis.  However, the SE field lines, which are
further away from the central source than the NW field lines, appear
to be roughly aligned with the jet axis.

\subsection{Comparison and Discussion}

As mentioned earlier, current polarization observations support a
gravitational collapse model of a rotating magnetized cloud core for
low-mass star formation.  For the first comparison, we compare to a simple
version of such model, in which the field lines are initially poloidal with
the field axis aligned with the rotation axis.  Because of the gravitational
collapse, the field lines are pinched toward the source, forming an
hourglass field morphology.  Moreover, toroidal fields are generated by
rotation in the flattened envelope near the source around the
rotation axis.  Figure \ref{fig:cont_pol}c shows the 3D view toward the
inner part of such a model from \citet{Kataoka2012}, where the toroidal
field lines have formed.  The model has a spatial scale about 4 times
larger because it is 5-10 times older than HH 211, and is thus only used as an
illustration of the complex structure of the magnetic field in the
envelope. The model has an edge-on view, appropriate for HH 211.  It is
rotated with the field axis (black lines) aligned with the jet axis in HH
211.  In addition, it has the same rotation sense as observed in HH 211,
with the blueshifted part in the SW and redshifted part in the NE
\citep{Lee2009}.

The observed field morphology is complicated, but may not be
incompatible with the model, as it predicts different orientation depending
on the region/distance.  The NW field lines can trace the toroidal
fields in the flattened envelope generated by the rotation.  The SE field
lines are further away from the center and thus can trace the fields at the
outer edge of the flattened envelope, where the fields are still mainly
poloidal.  The SW field lines are slightly pinched and thus can trace the
hourglass field lines dragged in by the gravitational collapse.  The NE
field lines are located closer to the source, where the hourglass field
lines are expected to be dragged in more by the gravitational collapse and
thus bent to be roughly perpendicular to the jet axis \cite[see also
Fig 4a in][]{Allen2003}.  Note that it is also possible that the field lines
there trace the toroidal fields generated by the rotation, as claimed in the
case of NGC 1333 IRAS 4A where field lines were also detected
perpendicular to the jet axis \citep{Crutcher2012}.  In
addition, the NE/SW asymmetry could also be partly due to the presence of a
secondary source. The polarization is not detected in all parts of the
envelope, probably because of depolarization due to the complicated field
morphology and our insufficient angular resolution \citep{Kataoka2012}. 
Further observations at higher sensitivity and resolution are needed
to confirm the different field orientations in different regions.

Hourglass field morphologies have been seen in other Class 0 systems,
e.g., NGC 1333 IRAS 4A (a binary system) at $\sim$ 400 AU resolution
\citep{Girart2006} and L 1157 at 300-525 AU resolution \citep{Stephens2013}.
At higher, $\sim$ 180 AU, resolution, toroidal field lines have also been
detected tentatively in NGC 1333 IRAS 4A within 300 AU of the center
\citep{Crutcher2012}.  In HH 211, a partial hourglass field morphology has
also been seen at $\sim$ 1000 AU resolution, with a pinched field morphology
in the SW \citep{Hull2013}.  Here at higher, $\sim$ 170 AU, resolution, we
may have detected not only the hourglass field morphology in the SW further
in but also the toroidal field lines in the flattened envelope within 300 AU
of the center.  Thus,
higher resolution observations are needed to reveal the toroidal fields,
likely because the rotation dominates only in the very inner region in the
Class 0 phase.

In current theory, the pinched geometry of the magnetic field in the
flattened envelope can produce a magnetic braking, preventing a RSD to be
formed at the center \citep{Allen2003}.  However, such a RSD must have
formed in order to launch the jet, as claimed in \citet{Lee2009}.  It has
been argued that a magnetic-field-rotation misalignment in a larger size
scale could help the formation of the disk \citep{Joos2012,Li2013}.  In the
larger size scale in the cloud scale (arc-minute scale)
\citep{Matthews2009}, the magnetic field lines are found to be roughly
north-south oriented, neither aligned with nor perpendicular to the jet
axis.  However, the rotation axis in this scale is found to be roughly
aligned with the jet axis \citep{Lee2009,Tanner2011}.  Thus, there is a
misalignment of $\sim$ 30\degree{} between the large-scale field axis and
the rotation axis, and it may help the disk formation.

\section{Jet}

\begin{figure} [!hbp]
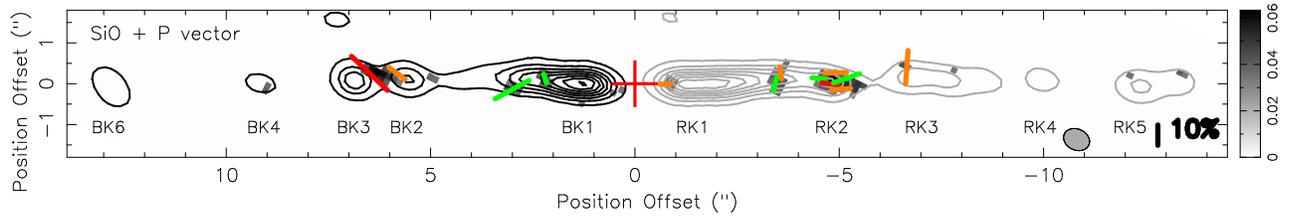

\centering
\putfig{0.68}{270}{f2.ps} %SiO_pol.ps
\figcaption[]
{HH 211 jet in SiO J=8-7, rotated clockwise to be aligned
with the x-axis. The black and gray contours show
the blueshifted and redshifted components of the jet, respectively.
The contours start from 4$\sigma$ with a step of 9$\sigma$, where $\sigma \sim 20$
\mJyb. The systemic velocity in HH 211 is $9.2\pm0.3$ \vkm{} LSR \citep{Lee2009}.  
The blueshifted component is obtained by averaging the emission
over the blueshifted velocity from
-14 to 2 \vkm{} and the redshifted component over the redshifted velocity
from 19 to 37 \vkm{}.
The gray-scale image shows the line polarization intensity greater than 
2$\sigma$ detection,
where the rms noise level $\sigma=18$ \mJyb{}.
The color and length of the polarization vectors have the same meanings as those in
Figure \ref{fig:cont_pol}.
 \label{fig:siojet} } \end{figure}

Figure \ref{fig:siojet} shows the blueshifted (black contours) and
redshifted (gray contours) components of the jet in SiO, rotated clockwise
to be aligned with the x-axis.  As seen in \citet{Lee2009}, the jet is
highly collimated and consists of a chain of knotty shocks.

\subsection{Line Polarization and Field Morphology}

The figure also shows the polarization intensity map and vectors in the jet
in the SiO line.  The angles of the polarization vectors have an
uncertainty of $\sim$ 13\degree{}. The line polarization can be attributed
to the GK effect.  In HH 211, the jet lies close to the plane of the sky. 
For the SiO J=8-7 emission, the optical depth is close to 1 and the
radiative transition rate is greater than the collision rate in most regions
\citep{Lee2009}.  All these are optimal for polarization detection and thus
the polarization degree could exceed 10\% \citep{Kylafis1983}, as seen in
the observation.  Two polarization detections, one in knot RK2 and one in
between knots BK2 and BK3, are greater than 3$\sigma$, and thus should be
real.  Their polarization vectors have different orientations, with the
former parallel to the jet axis and the latter inclined by $\sim$
50\degree{} to the jet axis.  Other detections are below 3$\sigma$ and thus
should be considered as tentative.  No polarization is detected toward the
bright innermost pair of knots probably because of the higher density there than
other regions.

According to the GK effect, the field could be either parallel or
perpendicular to the polarization vector, depending on the angle $\delta$
between the velocity flow axis and the polarization vector
\citep{Kylafis1983}.  If $\degreeqa{35}{3} < \delta <
\degreeqa{54}{7}$, the field direction is parallel to the polarization
vector.  If $0 < \delta < \degreeqa{35}{3}$, then there is a 90\degree{}
ambiguity in the field direction. In the jet, the velocity flow axis is the
same as the jet axis.  The polarization vectors in between knots BK2 and BK3
have $\delta \sim$ 50\degree{}, thus the field lines there are parallel to
the polarization vectors, highly inclined to the jet axis, and thus could be
helical.  In knot RK2, the polarization vectors are almost aligned with the
velocity flow axis.  In this case, the field lines could be either parallel
or perpendicular to the polarization vectors.  Therefore, the fields there
could be either poloidal or toroidal.

\subsection{Discussion}

In knot RK2, there are sufficient number of polarization vectors for
us to estimate the magnetic field strength there.  According to
\citet{Chandrasekhar1953} and \citet{Ostriker2001}, we can estimate the field strength in
the plane of the sky with the following
formula
\begin{equation} 
B \sim 0.5 \sqrt{4 \pi \rho} \frac{\triangle
v_\textrm{\scriptsize los}}{\triangle \phi} 
\end{equation} 
The mass density $\rho = 1.4 \nH2 \mH2 \sim 4.64\times10^{-16}$ g \cmc{} with 
$\nH2 \sim 10^7$ \cmc{}  \citep{Lee2009}. The
velocity dispersion along the line of sight $\triangle v_\textrm{\scriptsize
los} \sim 10$ \vkm{} \citep{Lee2009}.  With the dispersion of the
polarization angle $\triangle \phi \sim 20^\circ\pm13^\circ$ 
(including the uncertainty in the angle of the polarization vectors), 
we have $B \sim$ 35$^{+65}_{-14}$ mG. As discussed earlier, the field there could be toroidal.
According to \citet{Shu1995}, the toroidal field strength in an originally
unshocked jet material could be $\gtrsim$ 7 mG for the HH 211 jet, which
has a radius of $\lesssim$ 20 AU \citep{Lee2009}.
Thus, if the field in knot RK2 is purely toroidal,
our estimated field strength there is about 5 times as high,
probably not unreasonable considering a shock compression.
The magnetic pressure due to this toroidal field is
$B^2/8\pi \sim 3\times 10^{-5}$ dyne \cms{}.  The thermal pressure there is
$1.2 \nH2 k T \sim 8\times10^{-7}$ dyne \cms{} with $T \sim
500$ K \citep{Lee2009}, much lower than the magnetic pressure. 
Hence, the jet material there should be well confined by 
the magnetic field, if the field there is really toroidal.

Magnetic field morphology in protostellar jet is still poorly
determined due to a lack of polarization detection. Previously,
\citet{Carrasco2010} detected linear polarization in the synchrotron jet HH
80-81 from the high-mass protostar IRAS 18162-2048 and found that the field
lines there are mainly aligned with the jet axis.  They saw an increase in
polarization degree toward the jet edges and argued it to be due to helical
field toward the jet edges.  However, the jet with the polarization
detection is far at $\sim$ 0.5 pc away from the source and has an extremely
large transverse width of $\sim$ 40000 AU, and thus might not trace the
intrinsic jet coming from the source \citep{Esquivel2013}.  In comparison,
our polarization detections are within 2000 AU ($\sim$ \arcs{7}) of the
source.  The jet is located inside an outflow cavity and has a transverse
width of $\lesssim$ 40 AU \citep{Lee2009}, as expected for a jet coming from
a low-mass protostar.  Since the polarization detections here show
different orientations in different regions, further observations with ALMA
at higher sensitivity are really needed
to confirm them and to determine the field morphology. 

In conclusion, we might have detected for the first time polarized
SiO line emission in the jet due to the GK effect.  If confirmed, our
detection will open an perspective that GK effect can be used to infer
magnetic field morphology in the jet in the early phase of star formation,
in which the jet is primarily molecular.  The inferred field morphology can
then be used to constrain the jet launching models.

\acknowledgements
We thank the anonymous referee for insightful comments.
We thank the SMA staff for their efforts in running and maintaining the
array. C.-F.  Lee acknowledges grants from the
National Science Council of Taiwan (NSC 101-2119-M-001-002-MY3) and the
Academia Sinica (Career Development Award).

%% Remember to include "(" and ")" for the year,e.g., (1998)
%%

%The black curves outline the outflow cavity walls.  Cyan lines indicate the
%envelope plane.

\end{document}